\documentclass{sigkddExp}

\usepackage{color, colortbl}
\usepackage{changepage}
\usepackage{mathtools}
\usepackage{enumerate}
\usepackage{hyperref}
\usepackage{graphicx}
\usepackage{multirow}
\usepackage{amsmath}
\usepackage{tabularx}
\usepackage{booktabs}
\usepackage{flushend} 
\usepackage{amssymb}
\usepackage{float}
\usepackage{calc}
\usepackage{tcolorbox}

\tcbset{colback=pink}

\newcommand{\ignore}[1]{}
\DeclarePairedDelimiter\abs{\lvert}{\rvert}
\begin{document}


\title{Assessing Viewpoint Diversity in Search Results Using Ranking Fairness Metrics}

%

\numberofauthors{5}
%


\author{
%
\alignauthor Tim Draws \\
       \affaddr{Delft University of Technology}\\
       \affaddr{The Netherlands}\\
       \email{t.a.draws@tudelft.nl}
\alignauthor Nava Tintarev\\
       \affaddr{Delft University of Technology}\\
       \affaddr{The Netherlands}\\
       \email{n.tintarev@tudelft.nl}
\alignauthor Ujwal Gadiraju\\
       \affaddr{Delft University of Technology}\\
       \affaddr{The Netherlands}\\
       \email{u.k.gadiraju@tudelft.nl}
}
\additionalauthors{Additional authors: Alessandro Bozzon (Delft University of Technology, The Netherlands, email: {\texttt{a.bozzon@tudelft.nl}}) and Benjamin Timmermans
(IBM, The Netherlands, email: {\texttt{b.timmermans@nl.ibm.com}}).}
\date{30 July 1999}

\maketitle

\begin{tcolorbox}
This is an unedited manuscript accepted for publication in the \textit{ACM SIGKDD Explorations Newsletter}. The published version of the paper is available at \url{https://dl.acm.org/doi/10.1145/3468507.3468515}.
\end{tcolorbox}

\begin{abstract}
The way pages are ranked in search results influences whether the users of search engines are exposed to more homogeneous, or rather to more diverse viewpoints. However, this viewpoint diversity is not trivial to assess. In this paper, we use existing and novel ranking fairness metrics to evaluate viewpoint diversity in search result rankings. We conduct a controlled simulation study that shows how ranking fairness metrics can be used for viewpoint diversity, how their outcome should be interpreted, and which metric is most suitable depending on the situation. This paper lays out important groundwork for future research to measure and assess viewpoint diversity in real search result rankings.
\end{abstract}


\section{Introduction}

Search result rankings strongly influence user attitudes, preferences, and behavior \cite{Epstein2015,Joachims2005,Pan2007,Pogacar2017}. Underlying this effect are cognitive biases such as \emph{position bias}, which describes users' tendency to pay more attention to documents at higher ranks \cite{Joachims2005,Pan2007}. Recent research has demonstrated that these biases go to such an extent that rearranging search result rankings to favor different stances on the same topic can affect users' personal opinions \cite{Epstein2015,Pogacar2017}. To mitigate such unintentional biases, it is important to maintain a strong viewpoint diversity in search result rankings -- especially when they relate to disputed topics.

Viewpoint diversity in search result rankings is closely related to the notion of ranking fairness. The aim in fair ranking is to measure and adapt ranked lists in terms of their fairness concerning a given characteristic \cite{Biega2018,Sapiezynski2019,Yang2017}. For example, a ranked list of candidates on a job seeking platform could be evaluated with respect to gender fairness. A fair ranking is then considered to be one in which gender does not affect the ranking of candidates. Analogously in this paper, a search result ranking is evaluated with respect to \emph{viewpoint} -- to the best of our knowledge, a novel application of ranking fairness. Such a viewpoint can for example convey different stances on a topic, or different underlying reasons for a given stance. A search result ranking that is fair (or unbiased) with respect to viewpoints would give each viewpoint its fair share of coverage, contributing to viewpoint diversity in the search results.\footnote{Note that here we are thus looking at fairness in the \emph{outcome} of a ranking algorithm; i.e., not at procedural fairness.}

One major building block of studying viewpoint fairness in search result rankings is deciding how to measure it. Several metrics have been developed that assess fairness in rankings \cite{Yang2017,Sapiezynski2019} (see Section \ref{sec:relWork}). These metrics evaluate fairness in terms of \emph{statistical parity}, which is satisfied in a ranking if a given variable of interest -- here, the expressed viewpoint -- does not influence how documents are ranked. In this paper, we investigate whether ranking fairness metrics can be used to assess \textit{viewpoint diversity} in search results.

We generate a range of synthetic search result rankings with varying degrees of ranking bias and explore the behavior of existing and novel ranking fairness metrics on these rankings. For our use case of viewpoint diversity, we consider two fundamental scenarios: \emph{binomial viewpoint fairness}, in which the task is to measure viewpoint diversity with respect to one specific \emph{protected viewpoint}, and \emph{multinomial viewpoint fairness} where the aim is to protect all available viewpoints simultaneously. We make the following contributions:

\begin{enumerate}
    \item We present a simulation study that illustrates how existing ranking fairness metrics can be used to assess viewpoint diversity in search result rankings. We show how these metrics behave under varying conditions of viewpoint diversity and provide a guide for their use (Section \ref{sec:metricBehavior}).
    \item We propose a novel ranking fairness metric for assessing multinomial viewpoint fairness (Section \ref{sec:metrics}) and also analyze its behavior (Section \ref{sec:metricBehavior}).
\end{enumerate}

We find that all the considered ranking fairness metrics can distinguish well between different levels of viewpoint diversity in search results. However, which specific metric is most sensitive to a lack of viewpoint diversity depends on how many viewpoint categories there are, the distribution of advantaged and disadvantaged items in the ranking, and the severity of the ranking bias.

All code and supplementary material related to this research are openly available at \url{https://osf.io/nkj4g/}.

\begin{table*}
\centering
  \caption{The viewpoint label taxonomy we consider in this paper. Labels are denoted by $s$ and represented as ordinal values.}\label{tbl:viewpointCats}
  \begin{tabular}{rll}
    \toprule
    {\bfseries $s$} & {\bfseries Description} & {\bfseries Example}\\
    \midrule
    -3 & strongly opposing & ``Horrible places! All zoos should be closed ASAP.''\\
    -2 & opposing & ``We should strive towards closing all zoos.''\\
    -1 & somewhat opposing & ``Despite the benefits of zoos, overall I'm against them.''\\
    0 & neutral & ``These are the main arguments for and against Zoos.''\\
    +1 & somewhat supporting & ``Although zoos are not great, they benefit society.''\\
    +2 & supporting & ``I'm in favor of zoos, let's keep them.''\\
    +3 & strongly supporting & ``There is nothing wrong with zoos -- open more!''\\
  \bottomrule
\end{tabular}
\end{table*}

\section{Background and Related Work}\label{sec:relWork}

Diversity in search result rankings is not a novel topic. Several methods have been proposed to measure and improve diversity in ranked lists of search results~\cite{Abid2015,Agrawal2009,Clarke2008,Sakai2011,Sakai2019}. Unlike previous methods, which aim to \textit{balance} relevance (e.g., in relation to a user query) and diversity (e.g., in relation to user intent), we delve deeper into the notion of diversity. Specifically, we focus on ranking \emph{fairness} for assessing viewpoint diversity, which originates from the field of fair machine learning.

Fairness and the mitigation of bias in machine learning systems extends into several different sub-fields~\cite{Barocas2016,Bellamy2019,Ntoutsi2020}. One of these sub-fields -- \emph{fair ranking} -- has received increasing attention recently, following calls for dealing with bias on the web~\cite{Baeza-Yates2018}. This has led to the development of methods to increase ranking fairness~\cite{Biega2018,Celis2018,Singh2018a,Zehlike2017} as well as evaluative frameworks~\cite{Asudeh2019,Biega2018,Das2019,Sapiezynski2019} and metrics~\cite{Kulshrestha2019,Yang2017} for assessing bias and fairness in ranked lists. Measuring ranking fairness requires deciding which notion of fairness to handle (i.e., defining a \emph{fair ranking scenario})~\cite{Das2019,Biega2018} and discounting the metric computation by rank to account for differences in attention over the ranks~\cite{Biega2018,Sapiezynski2019}.

Previously proposed ranking fairness metrics commonly presuppose that a \emph{fair} or \emph{unbiased} ranking is one in which \emph{statistical parity} is present \cite{Yang2017}. A machine learning algorithm satisfies statistical parity when an item's probability of receiving a given outcome is not affected by belonging to a \emph{protected group} \cite{Verma2018}. Such a protected group can be any subset of the overall population of items that share some characteristic that is not supposed to affect the algorithm outcome. In the context of ranking, statistical parity holds when membership in a protected group has no influence on a document's position in the ranking \cite{Yang2017}. Suppose a user enters the query \emph{Should Zoos Exist?} into a web search engine, which then returns a ranked list of search results. Each document in the ranking corresponds to \emph{some} viewpoint concerning zoos or is neutral towards the topic. A ranking assessor could define the \emph{opposing} side of the zoo-argument as the protected viewpoint. Statistical parity would then be satisfied if expressing the protected viewpoint does not affect the ranking of documents.

Yang and Stoyanovich \cite{Yang2017} introduce three metrics that assess statistical parity in rankings. These metrics compare the group membership distribution (i.e., share of protected and non-protected items) in a ranking at different cut points (e.g., 10, 20, \ldots) with the ranking's overall group membership distribution. Aggregating the results of these comparisons in a discounted manner incorporates the intuition that an absence of bias is more important among higher ranks. We formalize the metrics introduced by Yang and Stoyanovich \cite{Yang2017} in Section \ref{sec:metrics}.

Ranking fairness has also been assessed in at least two other notable ways. First, Kulshrestha et al. \cite{Kulshrestha2019} introduce a metric that quantifies ranking bias related to continuous attributes as opposed to group membership. Their metric considers the mean of a continuous variable of interest at each step of its computation. Despite this promising groundwork, measuring continuous ranking bias remains limited; for instance, by considering only the mean, other important characteristics of continuous distributions (i.e., such as the standard deviation or distribution type) are ignored. Second, recent work has defined \emph{criteria} that a ranking has to fulfill to be considered \emph{fair}~\cite{Zehlike2017,Sapiezynski2019}. Whether the ranking fulfills these criteria is assessed using null hypothesis significance testing. Our aim, however, is to quantify the \emph{degree} of viewpoint diversity in search result rankings.

\section{Measuring Fairness in Rankings}\label{sec:measuringFairness}

Viewpoint diversity in search results can best be illustrated by a running example. Consider that a user wants to form an educated opinion on the topic `\emph{Should Zoos Exist?}', and turns to web search to gather information. Let us assume that each document that the user encounters in the search result list will express a viewpoint concerning zoos or be neutral towards the topic.\footnote{Here, \emph{neutral} could mean that a document is not opinionated, provides a balanced overview of the different viewpoints, or is irrelevant to the topic.}  These viewpoints can be represented in an ordinal manner, as illustrated in Table \ref{tbl:viewpointCats}. We thus categorize the different viewpoints related to zoos by placing them on a 7-point scale ranging from \emph{strongly opposing} to \emph{strongly supporting} the existence of zoos.\footnote{Note that this is just one possible way to categorize existing viewpoints on a topic.}

\subsection{Preliminaries and Notation}\label{sec:prelNot}

We are given a set of documents $D$ and a set of viewpoint labels $S$. Both sets contain the same number of elements $N$. Each document $d \in D$ is uniquely associated with one label $s_d \in S$. Here, $s_d$ reflects the viewpoint of document $d$ towards a given disputed topic, rated on a 7-point scale ranging from \emph{extremely opposing} to \emph{extremely supporting}. The viewpoint labels in $S$ are integers ranging from $-3$ to $3$, where negative values indicate an \emph{opposing viewpoint}, 0 indicates a \emph{neutral viewpoint}, and positive values indicate a \emph{supporting viewpoint} towards the debated topic (see Table \ref{tbl:viewpointCats} for an example). A ranked list of $D$ is denoted as $\tau$. We denote the number of items that belong to a subset $p$ of $S$ as $S^{p}$, which becomes $S^{p}_{1\ldots i}$ when constrained to the top $i$ ranked documents. Table \ref{tbl:notation} presents an overview of the notation introduced here.

\begin{table}
\centering
  \caption{Notation used throughout this paper.}\label{tbl:notation}
  \scalebox{0.9}{
  \begin{tabular}{ll}
    \toprule
    {\bfseries Notation} & {\bfseries Description}\\
    \midrule
    $d$ & document\\
    $D$ & set of documents\\
    $s_d$ & viewpoint label of document $d$\\
    $S$ & set of viewpoint labels\\
    $S^{p}$ & number of items in set $S$ that belong to subset $p$\\
    $\tau$ & ranked list of set $D$\\
    $S^{p}_{1\ldots i}$ & $S^{p}$ in the top $i$ ranked documents\\
    $N$ & number of elements in $D$, $S$, and $\tau$\\
  \bottomrule
\end{tabular}}
\end{table}

\subsection{Defining Fairness and Viewpoint Diversity}\label{sec:fairnessNotions}

There are many definitions of fairness, and so before describing fairness metrics, we first identify which type of fairness to handle. In this paper, we focus on the notion of \emph{statistical parity} (also commonly referred to as \emph{group fairness}; see Section \ref{sec:relWork}). This notion allows us to define several fairness aims for assessing viewpoint diversity. We consider two such aims, which we call \emph{binomial viewpoint fairness} and \emph{multinomial viewpoint fairness}. Below we describe these aims and align them with the notion of statistical parity in rankings.

\paragraph*{\textbf{Binomial viewpoint fairness}}
One aim for viewpoint diversity may be to treat one specific viewpoint, e.g., a minority viewpoint, fairly. For example, if a search result ranking on the query \emph{Should Zoos Exist?} is dominated by arguments \emph{supporting} zoos, the ranking assessor may want to evaluate whether the minority viewpoint (i.e., \emph{opposing} zoos) gets its fair share of coverage. The assessor may consider a binomial classification of documents into one of two groups: expressing the minority viewpoint or not expressing the minority viewpoint. Here, expressing the minority viewpoint is analogous to a protected group. Statistical parity in a ranking of such documents is satisfied when expressing the minority viewpoint does not affect a document's position in the ranking.

\paragraph*{\textbf{Multinomial viewpoint fairness}}
Another aim when evaluating viewpoint diversity may be that \emph{all} viewpoints are covered fairly. For example, a search result ranking on the query \emph{Should Zoos Exist?} could be assessed without explicitly defining a specific viewpoint as the protected group but instead considering the distribution over several existing viewpoints. Here the assessor thus considers a multinomial classification of documents into some viewpoint taxonomy (e.g., into seven categories depending on polarity and severity of the viewpoint; see Section \ref{sec:prelNot}). In this case, we say that statistical parity is satisfied when for each viewpoint, the choice of viewpoint does not influence a document's position in the ranking. Multinomial viewpoint fairness is thus more fine-grained than binomial viewpoint fairness: whereas binomial viewpoint fairness focuses on fairness towards one protected viewpoint, multinomial viewpoint fairness requires being fair to all viewpoints simultaneously.

\subsection{Desiderata and Practical Considerations for Metrics}\label{sec:pracCons}

\paragraph*{\textbf{Evaluating statistical parity}}

In this paper, we use ranking fairness metrics to assess viewpoint diversity in search result rankings. These are based on the notion of statistical parity, which is present in a ranking when the viewpoints that documents express do not affect their position in the ranking. However, we are only given the ranking and viewpoint per document and cannot assess the ranking algorithm directly. Statistical parity thus needs to be approximated.

We choose to approximate statistical parity in the same way as previously developed ranking fairness metrics \cite{Yang2017}. These metrics measure the extent to which the document distribution over groups (e.g., the protected and non-protected group) is the same in different top-$i$ portions of the ranking compared to the overall ranking (see Section \ref{sec:relWork}). The more dissimilar the distribution at different top-$i$ is from the overall distribution, the less fair the ranking.

\paragraph*{\textbf{Discounting the ranking fairness computation}}

User attention depletes rapidly as the ranks go up \cite{Joachims2005,Pan2007}. For example, in a regular web search, the majority of users may not even view more than 10 documents. This means that a measure of viewpoint diversity needs to consider the rank of documents, and not just whether viewpoints are present. More specifically, fairness is more important at higher ranks.

A practical way to incorporate this notion into a ranking fairness metric is to include a discount factor. Sapiezynski et al. \cite{Sapiezynski2019} point out that such a discount depends on the user model related to the particular ranking one is assessing. Similar to the ranking fairness metrics introduced by Yang and Stoyanovich \cite{Yang2017}, we choose the commonly used $\text{log}_2$ discount for each metric we introduce below. Yang and Stoyanovich \cite{Yang2017} suggest discounting in steps of 10 (see Section \ref{sec:relWork}). Such a binned discount nicely incorporates the notion that ranking fairness is more important in the top 10 documents than it is in the top 20 documents. However, especially on the first page of search results, individual ranks matter a lot \cite{Joachims2005,Pan2007}. 
We therefore decide to discount by individual rank and consider the top 1, 2, ... $N$ documents at each step of the aggregation.

\paragraph*{\textbf{Normalization}}

When evaluating and comparing metrics, it is useful if they all operate on the same scale. We thus only consider normalized ranking fairness metrics.

\subsection{Ranking Fairness Metrics}\label{sec:metrics}

In this section, we describe the metrics that we use to assess viewpoint diversity in search result rankings. These metrics are partly based on existing ranking fairness metrics and partly novel. We adapt each metric that we use to fit the practical considerations outlined in Section \ref{sec:pracCons}. Taking these practical considerations into account, we define a template that each normalized ranking bias (nRB) metric that we use will follow:

\begin{equation}
\label{eq:template}
    \text{nRB}(\tau) = \frac{1}{Z} \sum_{i=1}^{N} \frac{F(i)}{\text{log}_{2}(i+1)}.
\end{equation}

\noindent Here, $F$ is a function that quantifies the ranking bias in the ranked list $\tau$. All metrics that we describe in the following subsections will only differ in terms of how they define $F$. The function $F$ is iteratively computed for the top $i$ documents and subsequently aggregated by using a $\text{log}_2$ discount. Finally, $Z$ is a normalizing constant that takes on the value for $F$ given the maximally unfair permutation of $\tau$.\footnote{A description of how we normalize each metric can be found at \url{https://osf.io/nkj4g/}.}

\subsubsection{Metrics to assess binomial viewpoint fairness}

Yang and Stoyanovich \cite{Yang2017} propose three ranking fairness metrics to assess statistical parity in rankings (see Section \ref{sec:relWork}). We interpret these metrics to fit binomial viewpoint fairness and adapt them to fit the considerations outlined in Section \ref{sec:pracCons}.

Note that, although we define a protected and a non-protected viewpoint before using any of these metrics, the metrics are in principle agnostic as to which of the two viewpoint categories (i.e., ``protected'' and ``unprotected'') is advantaged in the ranking. That is, they do not only measure when the protected viewpoint is treated unfairly but also capture if a ranking is biased \emph{towards} the protected viewpoint. The categorization into protected and non-protected viewpoints should thus be viewed as a binary classification of documents that -- in a fair scenario -- does not affect how documents are ranked.

\paragraph*{\textbf{Normalized Discounted Difference (nDD)}} This metric computes the difference between the proportion of items that belong to the protected group at different top-$i$ subsets of the ranking with and overall proportion:

\begin{equation}
    \text{nDD}(\tau) = \frac{1}{Z} \sum_{i=1}^{N} \frac{1}{\text{log}_{2}(i+1)} \abs*{\frac{S^{p}_{1\ldots i}}{i}-\frac{S^{p}}{N}}.
\end{equation}

\noindent Here, $S^{p}$ is the number of documents in the protected group and $N$ is the total number of ranked documents. 

\paragraph*{\textbf{Normalized Discounted Ratio (nDR)}} This metric measures the difference between the ratio of documents that express the protected viewpoint, and those who do not, at different top-$i$ portions of the ranking with the overall ratio:

\begin{equation}
    \text{nDR}(\tau) = \frac{1}{Z} \sum_{i=1}^{N} \frac{1}{\text{log}_{2}(i+1)} \abs*{\frac{S^{p}_{1\ldots i}}{S^{u}_{1\ldots i}}-\frac{S^{p}}{S^{u}}}.
\end{equation}

\noindent Here, $S^{u}$ refers to the number of documents that do not express the protected viewpoint. Here we set the value of fractions to 0 if their denominator is 0 \cite{Yang2017}.

\paragraph*{\textbf{Normalized Discounted Kullback-Leibler Divergence (nDKL)}} This metric makes use of the \textit{Kullback-Leibler divergence} (KLD), an asymmetric measure of difference between probability distributions \cite{kullback1951information}. For two discrete probability distributions $P$ and $Q$ that are defined on the same probability space $\mathcal{X}$, KLD is given by

\begin{equation}
    \text{KLD}(P||Q) = \sum_{x \in \mathcal{X}} P(x) \log \Bigg( \frac{P(x)}{Q(x)} \Bigg).
\end{equation}
    
\noindent To measure binomial viewpoint fairness in a ranking, $P$ and $Q$ can be defined as

\begin{align*}
    P = \Big( \frac{S^{p}_{1\ldots i}}{i}, \frac{S^{u}_{1\ldots i}}{i} \Big), Q = \Big( \frac{S^{p}}{N}, \frac{S^{u}}{N} \Big).
\end{align*}

\noindent This way, KLD measures the divergence between the proportion of protected items at rank $i$ and in the ranking overall.\footnote{Note that KLD is not defined for $P = (0,1)$. In this case, we smooth to $P = (0.001,0.999)$.} We can insert KLD in Equation \ref{eq:template}:

\begin{equation}
    \text{nDKL}(\tau) = \frac{1}{Z} \sum_{i=1}^{N} \frac{\text{KLD}(P|| Q)}{\text{log}_{2}(i+1)}. 
\end{equation}

\subsubsection{Metric to assess multinomial viewpoint fairness}

To the best of our knowledge, no metrics have so far been proposed that explicitly assess ranking fairness for multiple categories at once. The previously introduced nDKL metric can in principle be expanded to assess multinomial viewpoint fairness. KLD measures the distance between two discrete probability distributions $P$ and $Q$. In the multinomial case, we can define $P$ and $Q$ as multinomial distributions over the available viewpoint categories. For instance, in our use case of viewpoints rated on a 7-point scale, $P$ and $Q$ may be given by:

\begin{align*}
    P &= \Big( \frac{S^{-3}_{1\ldots i}}{i}, \frac{S^{-2}_{1\ldots i}}{i}, \frac{S^{-1}_{1\ldots i}}{i}, \frac{S^{0}_{1\ldots i}}{i}, \frac{S^{+1}_{1\ldots i}}{i}, \frac{S^{+2}_{1\ldots i}}{i}, \frac{S^{+3}_{1\ldots i}}{i} \Big),\\ 
    Q &= \Big( \frac{S^{-3}}{N}, \frac{S^{-2}}{N}, \frac{S^{-1}}{N}, \frac{S^{0}}{N}, \frac{S^{+1}}{N}, \frac{S^{+2}}{N}, \frac{S^{+3}}{N} \Big),
\end{align*}

\noindent where $S^{-3, -2, ..., 3}$ refer to the number of items in each viewpoint category. 

A problem with using KLD for multinomial distributions is that its normalization becomes extremely complex. To normalize KLD, the maximally divergent distribution of items needs to be computed at each step. Whereas this is rather straightforward in the binomial case,\footnote{A description of how we normalize each metric can be found at \url{https://osf.io/nkj4g/}.} finding the maximally divergent distribution becomes extremely expensive when more categories are added.

To resolve the normalization issue that comes with KLD, we propose a new metric that uses the Jensen-Shannon divergence (JSD) as an alternative distance function. Similarly to KLD, JSD measures the distance between two discrete probability distributions $P$ and $Q$ that are defined on the same sample space $\mathcal{X}$ \cite{Fuglede2004}. JSD can in fact be expressed using KLD:

\begin{equation*}
    \text{JSD}(P||Q) = 0.5 * \Big( KLD(P||R) + KLD(Q||R) \Big).
\end{equation*}

\noindent Here, $R = 0.5 * (P+Q)$ is the mid-point between $P$ and $Q$. In contrast to KLD (which can go to infinity), JSD is bound by 1 as long as one uses a base 2 logarithm in its computation \cite{Lin1991}. Knowing this maximally possible value for JSD, also an aggregated, discounted version of JSD is easily normalized. We thus propose \emph{Normalized Discounted Jensen-Shannon Divergence} (nDJS) as given by

\begin{equation}
    \text{nDJS}(\tau) = \frac{1}{Z} \sum_{i=1}^{N} \frac{\text{JSD}(P|| Q)}{\text{log}_{2}(i+1)},
\end{equation}

\noindent where $\text{JSD}(P|| Q)$ is the JSD between $P$ and $Q$. Although we here propose nDJS specifically for assessing multinomial viewpoint fairness, note that it can be used to assess binomial viewpoint fairness as well.

\section{Simulation Study}\label{sec:simulation}

\begin{table*}
\centering
\caption{Examples of sample weight allocations for the simulation of binomial (left-hand table, $\alpha = 0.5$) and multinomial viewpoint fairness (right-hand table; $\alpha = -0.8$).}\label{tbl:sampleWeights}
\begin{tabular}{lccccccc}
    \toprule
    {\bfseries viewpoint} & -3 & -2 & -1 & 0 & +1 & +2 & +3\\
    \midrule
     {\bfseries weight} & $w_1$ & $w_1$ & $w_1$ & $w_2$ & $w_2$ & $w_2$ & $w_2$\\
     {\bfseries (rounded)} & 0.5 & 0.5 & 0.5 & 1.5 & 1.5 & 1.5 & 1.5\\
  \bottomrule
\end{tabular}
\quad
\begin{tabular}{lccccccc}
    \toprule
    {\bfseries viewpoint} & -3 & -2 & -1 & 0 & +1 & +2 & +3\\
    \midrule
     {\bfseries weight} & $w_{2}$ & $w_{2}$ & $w_{1}$ & $w_{2}$ & $w_{2}$ & $w_{2}$ & $w_{2}$\\
     {\bfseries (rounded)} & 0.2 & 0.2 & 1.8 & 0.2 & 0.2 & 0.2 & 0.2\\
  \bottomrule
\end{tabular}
\end{table*}

In this section, we show how the metrics introduced in Section \ref{sec:metrics} behave in different ranking scenarios. Our code to implement the metrics and simulation is openly available.\footnote{See our repository at \url{https://osf.io/nkj4g/}.}

\subsection{Generating Synthetic Rankings}\label{sec:rankingGen}

To simulate different ranking scenarios, we first generate three synthetic sets $S1$, $S2$, and $S3$ to represent different viewpoint distributions. The items in each set simulate viewpoint labels for 700 documents (i.e., to enable a simple balanced distribution over seven viewpoints) and are distributed as shown in Table \ref{tbl:Sdists}. Whereas $S1$ has a balanced distribution of viewpoints, $S2$ and $S3$ are skewed towards supporting viewpoints.\footnote{Due to symmetry we do not include similar distributions for opposing viewpoints.} We use $S1$, $S2$, and $S3$ to simulate both binomial and multinomial viewpoint fairness.\footnote{Because we are only interested in rankings with respect to viewpoint labels, we do not generate any actual documents here. Instead, we rank the labels themselves.}

\begin{table}[ht]
\centering
\caption{Viewpoint distributions of the sets $S1$, $S2$, and $S3$.}\label{tbl:Sdists}
\begin{tabular}{lccccccc}
    \toprule
     & \textbf{-3} & \textbf{-2} & \textbf{-1} & \textbf{0} & \textbf{+1} & \textbf{+2} & \textbf{+3}\\
    \midrule
    {$S1$} & 100 & 100 & 100 & 100 & 100 & 100 & 100\\
    {$S2$} & 80 & 80 & 80 & 115 & 115 & 115 & 115\\
    {$S3$} & 60 & 60 & 60 & 130 & 130 & 130 & 130\\
  \bottomrule
\end{tabular}
\end{table}

\paragraph*{\textbf{Sampling}} We create rankings of the viewpoint labels in $S1$, $S2$, and $S3$ by conducting a weighted sampling procedure. To create a ranking, viewpoint labels are gradually sampled from one of the three sets without replacement to fill the individual ranks. Each viewpoint label in the set is assigned one of two different sample weights that determine the labels' probability of being drawn. These two sample weights are controlled by the ranking bias parameter $\alpha$ and given by: $w_{1} = 1.0001 - 1 \times \alpha$; $w_{2} = 1.0001 + 1 \times \alpha$.

\paragraph*{\textbf{Alpha}} For our simulation of binomial and multinomial viewpoint fairness, ranking bias is controlled by the continuous parameter $\alpha$ = [$-1$, $1$]. More specifically, $\alpha$ controls the sample weights $w_1$ and $w_2$ that are used to create the rankings. Whereas a negative $\alpha$ will result in higher ranks for viewpoints that are assigned $w_1$, a positive $\alpha$ will advantage viewpoints that are assigned $w_2$. The further away $\alpha$ is from 0, the more extreme the ranking bias. If $\alpha$ is set to exactly 0, no ranking bias is present: here it does not matter whether a viewpoint label is assigned $w_1$ or $w_2$, the sample weights are the same. In each simulation, we try 21 degrees of ranking bias for $\alpha = -1$ to $\alpha = 1$ in steps of $0.1$.

\subsubsection{Simulating binomial viewpoint fairness}

To simulate binomial viewpoint fairness, we create ranked lists from $S1$, $S2$, and $S3$ with different degrees of ranking bias. Ranking bias -- controlled by $\alpha$ -- in this scenario refers to the degree to which expressing a protected viewpoint influences a document's position in the ranking. We define all \emph{opposing} viewpoints (i.e., -3, -2, and -1) together as the protected viewpoint and assign them the sample weight $w_1$. All other viewpoints (i.e., 0, 1, 2, and 3) are thus non-protected and assigned the other sample weight $w_2$ when generating the rankings. Table \ref{tbl:sampleWeights} (left-hand table) shows an example of this sample weight allocation for $\alpha = 0.5$. In this example, the non-protected viewpoint is more likely to be drawn compared to the protected viewpoint.

Our weighted sampling procedure (see above) will produce slightly different rankings even when the same $\alpha$ is used. To get reliable results, we therefore create 1000 ranked lists for each $\alpha$ and aggregate the results.

\subsubsection{Simulating multinomial viewpoint fairness}

We simulate multinomial viewpoint fairness by again sampling rankings from $S1$, $S2$, and $S3$ with different degrees of ranking bias. This time the ranking bias $\alpha$ is defined as how much the expressed viewpoint generally affects a document's position in the ranking.

Since there are many scenarios in which one (or more) of several viewpoint categories could be preferred over others in a ranking, we focus on just one specific case: our simulation prefers \emph{one} of the seven viewpoints over the other six. For example, this could be the case if a search result list is biased towards an extremely opposing viewpoint. We randomly assign the sample weight $w_1$ to one of the opposing viewpoints (i.e., $-3$, $-2$, or $-1$) and the sample weight $w_2$ to all remaining viewpoints for each ranking we create. This means that each ranked list we create prefers a different viewpoint, reflecting the idea that we do not know which viewpoint might be preferred before evaluating the ranking and we have no specific, pre-defined protected viewpoint. Table \ref{tbl:sampleWeights} (right-hand table) shows an example of this sample weight allocation for $\alpha = -0.8$. In this example, the ranked list will prefer the viewpoint $-1$ over all other viewpoints. We again compute 1000 ranked lists for each $\alpha$ and aggregate the results.

\begin{figure*}
\centering
\includegraphics[width=\textwidth]{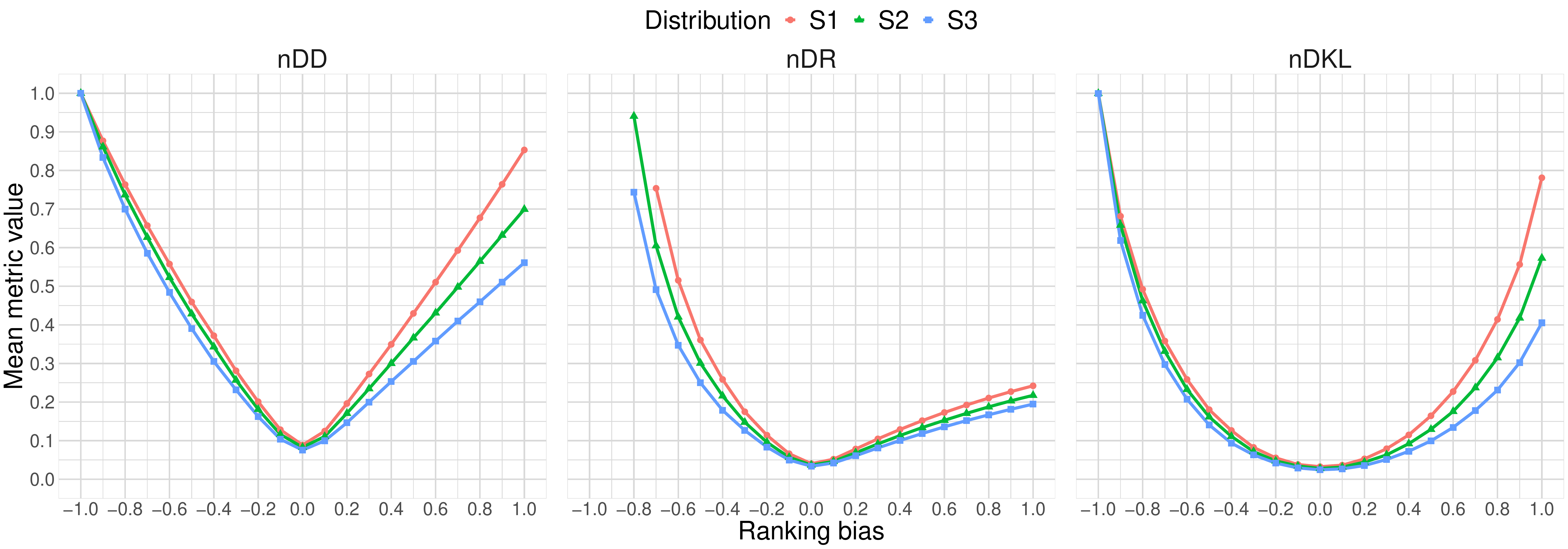}
\caption{Behavior of the metrics nDD (left-hand plot), nDR (center plot), and nDKL (right-hand plot) on the sets $S1$ ($S^p = 300$), $S2$ ($S^p = 240$), and $S3$ ($S^p = 180$) across different $\alpha$ (ranking bias) settings.
} \label{fig:metricPerf}
\end{figure*}

\subsection{Metric Behavior}\label{sec:metricBehavior}

Here, we explore the behavior of the ranking fairness metrics introduced in Section \ref{sec:metrics} using the synthetic rankings from Section \ref{sec:rankingGen}.

\subsubsection{Binomial viewpoint fairness}

Binomial viewpoint fairness can be assessed using nDD, nDR, or nDKL. Each of these metrics measures the degree to which expressing a protected viewpoint affects the ranking of documents. 
The ranking in our running example is considered fair if documents opposing zoos (i.e., $-3$, $-2$, and $-1$) get a similar coverage throughout the ranking compared to other viewpoints (i.e., $0$, $+1$, $+2$, and $+3$). A fair scenario should lead to a low score on each of the three metrics.

Figure \ref{fig:metricPerf} shows the mean outcome of nDD, nDR, and nDKL from 1000 ranked lists per data set (i.e., $S1$, $S2$, and $S3$) and $\alpha$ (i.e., ranking bias) setting. Each set represents a different overall distribution of viewpoints (see Table \ref{tbl:Sdists}).

We note three characteristics that all three metrics share. First, each of the three metrics is lowest for low bias ($\alpha = 0$) and increases from there as the absolute value of $\alpha$ increases. This means that all three metrics function as expected: they produce higher values as ranking bias becomes more extreme. Second, each metric shows a steeper curve as the data sets contained fewer items that express the minority viewpoint (here, the protected opposing viewpoint) increases; i.e, $S1 > S2 > S3$. Different levels of ranking bias thus become easier to detect when the distribution of protected and non-protected items is more balanced. Third, each metric produces higher values for $\alpha = -1$ (protected viewpoint is \emph{advantaged}) than for $\alpha = 1$ (protected viewpoint is \emph{disadvantaged}). The reason behind this is that unfair treatment becomes increasingly harder to detect as the number of items in the disadvantaged group shrinks: if one group only encompasses around 25\% of items (e.g., such as in $S3$), it is less odd to see several items of the other group ranked first than if the distribution is more balanced. That is also why each metric produces higher values at $\alpha = 1$ as the number of protected items increases. 

Next to these general characteristics that are shared by all metrics, below we discuss differences that distinguish the metrics in terms of their behavior.

\paragraph*{\textbf{Normalized Discounted Difference}} For each of the three data sets, nDD reaches its maximum value of 1 when $\alpha = -1$ and is at its lowest with mean values of approximately 0.08 when $\alpha = 0$. Depending on the number of items that express the protected viewpoint, nDD reaches mean values between 0.55 and 0.85 when $\alpha = 1$ for the three data sets in our simulation. The curves for nDD in Figure \ref{fig:metricPerf} are also comparatively steep. This indicates that nDD is especially useful for distinguishing low levels of ranking bias.

\paragraph*{\textbf{Normalized Discounted Ratio}} The lowest mean nDR values in our simulation (reached at $\alpha = 0$ for each of the three data sets) all approximate 0.04. Even more so than nDD, nDR reaches mean values far below 1 when the protected viewpoint is disadvantaged in the ranking. The mean values for this form of extreme ranking unfairness range from approximately 0.19 to 0.24 in our simulation, depending on the number of protected viewpoint items. In comparison to the other two metrics, nDR is less steep than nDD but steeper than nDKL. It could thus be useful for detecting medium levels of ranking bias. However, if a ranking is unfair towards the minority viewpoint, nDR does not distinguish different levels of ranking bias well. We also find that our normalization procedure (i.e., dividing each metric outcome by the outcome for a maximally unfair ranking) does not normalize nDR correctly. Thus, the maximal mean values for nDR (which it reaches at $\alpha = -1$) lie above 1 and are therefore not displayed in Figure \ref{fig:metricPerf} (which has 1 as its upper limit).\footnote{We explore the reason behind this (including an alternative way to normalize nDR) in a supplementary document on our normalization procedures. This document can be found at \url{https://osf.io/nkj4g/}.}

\paragraph*{\textbf{Normalized Kullback-Leibler Divergence}} Similar to the other metrics, nDKL reaches its maximum value of 1 at $\alpha = -1$. In our simulation, the lowest mean values for nDKL (reached at $\alpha = 0$) approximated 0.03. Extremely positive $\alpha$ settings (i.e., disadvantaging the minority viewpoint) produce mean nDKL values between 0.40 and 0.78, depending on the number of items that express the minority viewpoint. Furthermore, nDKL has a more parabolic shape compared to nDD and nDR. Whereas nDKL can thus not distinguish low values of ranking bias well, it is useful for differentiating between high levels of ranking bias.

\subsubsection{Multinomial viewpoint fairness}

To assess multinomial viewpoint fairness, we use nDJS. This metric measures the degree to which the viewpoint that documents express is a factor for a ranking in general. For example, in a search result ranking related to the topic \emph{Should Zoos Exist?}, a range of viewpoints may exist, some of which may be advantaged in the ranking over other viewpoints. That is why we cannot use binomial ranking fairness metrics here: we do not have a specific viewpoint to protect, but instead wish to protect all viewpoints equally. A maximally fair ranking scenario would give all viewpoints a coverage across the ranking that is proportional to their share in the overall distribution. For (approximately) fair rankings, nDJS should return a low value.

\begin{figure}
\centering
\includegraphics[width=0.33\textwidth]{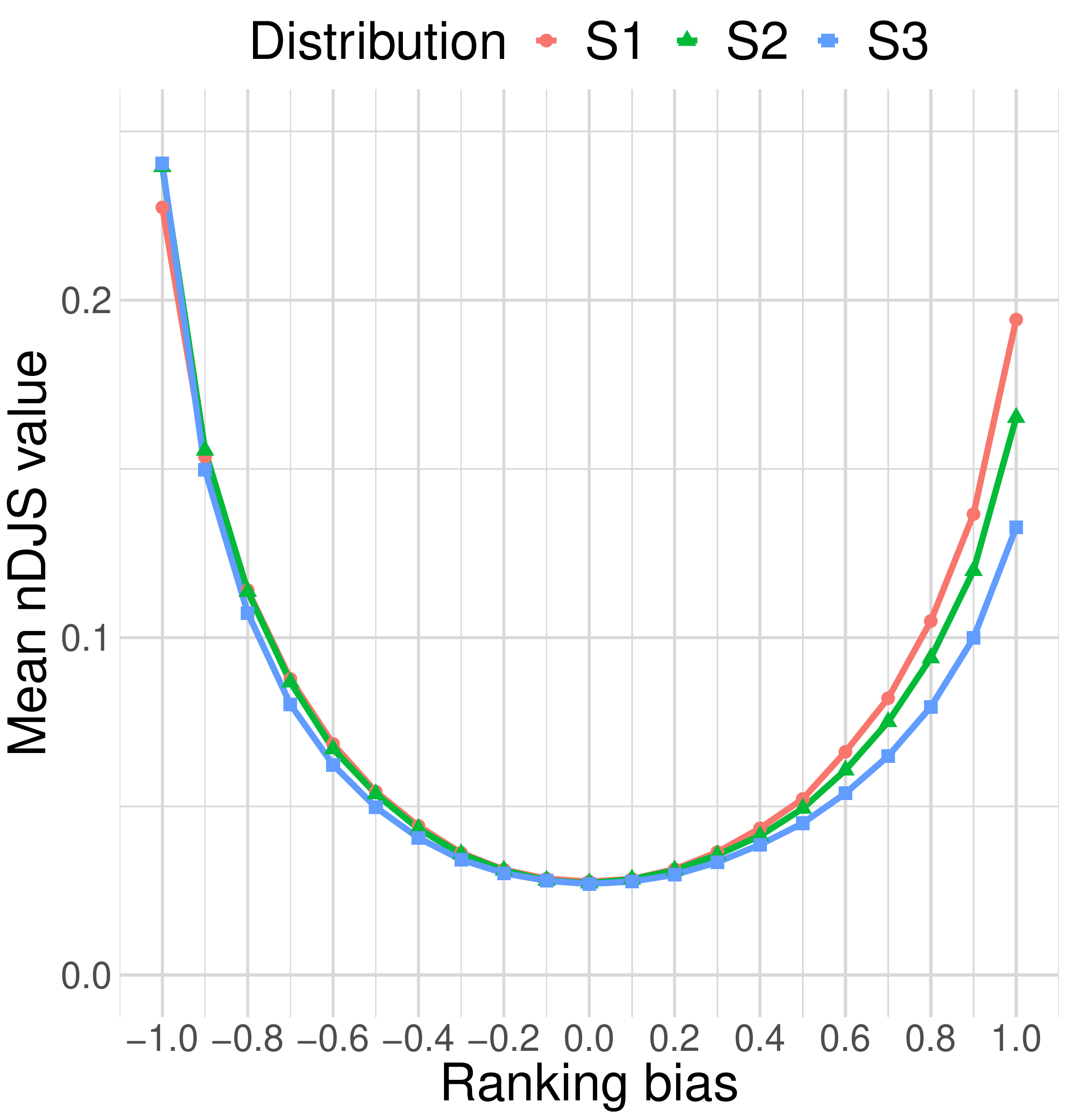}
\caption{Behavior of nDJS on the sets $S1$, $S2$, and $S3$ across different $\alpha$ (ranking bias) settings. The number of items with sample weight $w_1$ for rankings from the sets $S1$, $S2$, and $S3$ are 100, 80, and 60, respectively.
} \label{fig:metricPerfnDJS}
\end{figure}

We test nDJS on synthetic rankings that simulate varying degrees of bias on three different sets of items ($S1$, $S2$, and $S3$, see Section \ref{sec:rankingGen}). Figure \ref{fig:metricPerfnDJS} shows the mean outcome of nDJS from 1000 ranked lists per set and $\alpha$ (i.e., ranking bias) setting. Similar to the binomial ranking fairness metrics, nDJS does what it is expected to do: it produces its highest values at extreme $\alpha$ (ranking bias) settings and its lowest values at $\alpha = 0$. This means that nDJS can pick up the nuanced multinomial viewpoint fairness in our synthetic rankings. We observe, however, that due to its normalization, the maximum values for nDJS are much lower than for the metrics that assess binomial viewpoint fairness. When $\alpha = -1$ (i.e., when one random viewpoint is \emph{disadvantaged} compared to others), nDJS produces mean values between approximately 0.18 and 0.21. Due to the different normalization, it is therefore not possible to compare results from nDJS directly to results from the binomial ranking fairness metrics. For low values of ranking bias, the mean nDJS values approximate 0.03 on all three data sets. The mean nDJS value lies between approximately 0.07 and 0.09 when $\alpha = 1$ (i.e., when one viewpoint is \emph{advantaged} compared to others).

Similar to the binomial fairness metrics, the values that nDJS produces is again influenced by the proportion of advantaged items in the ranking. The more balanced this ratio, the easier it is to detect a ranking bias (i.e., the higher nDJS). Note that in this simulation, the distribution of advantaged and disadvantaged items was far from balanced, as we only treated one viewpoint label differently per ranking.

\newpage

\section{Discussion} 

In this section, we summarize our findings, provide a guide to using the metrics we examined, and discuss the limitations and implications of this research.

\subsection{Binomial Viewpoint Fairness}

Each of the three metrics we tested in our simulation can measure binomial viewpoint fairness (nDD, nDR, nDKL; see Section \ref{sec:metricBehavior}). However, depending on the distribution of protected and non-protected items, as well as the direction and level of ranking bias, a different metric might be suitable. Table \ref{tbl:metricRecs} shows which metric we recommend using in which scenario. In sum, we suggest taking the following considerations when assessing binomial viewpoint fairness:

\begin{enumerate}
    \item Generally, the more balanced the overall distribution of protected and non-protected items in the ranking, the better the metrics are able to distinguish different levels of ranking bias. When ranking bias is disadvantaging a protected group that only contains a small number of items, nDR appears to be the most suitable metric because it is the least vulnerable in this case.
    \item Which metric is most suitable also depends on how severe the bias in the ranking is estimated to be. Whereas nDD outputs the most divergent values for mild cases of ranking bias, nDKL distinguishes more severe cases of ranking bias better. Although nDR is slightly better in distinguishing medium levels of negative ranking bias, we do not recommend using it at all due to its normalization issues and weak performance when ranking bias is positive.
    \item If the minority viewpoint is preferred in the ranking, ranking bias is well detected by all three metrics. However, when the minority group is \emph{disadvantaged}, all metrics show a decrease in performance. In this case, we suggest using either nDD or nDKL, depending on how strong the ranking bias is.
\end{enumerate}

\begin{table}
\centering
\caption{Recommended metrics for different scenarios of ranking bias and overall viewpoint distribution (i.e., protected and non-protected items) in a ranked list.}\label{tbl:metricRecs}
  \begin{tabular}{llccc}
    \toprule
    & & \multicolumn{3}{c}{\bfseries Ranking Bias} \\
    & & Low & Medium & High \\
    \midrule
    \multirow{3}{*}{\bfseries Distribution} & Low balance & nDD & nDD & nDD \\
     & Medium balance & nDD & nDD & nDKL \\
     & High balance & nDD & nDKL & nDKL \\
    \bottomrule
  \end{tabular}
\end{table}

\subsection{Multinomial Viewpoint Fairness}

We find that our novel metric nDJS can assess multinomial ranking fairness. Similarly to the binomial fairness metrics, nDJS can distinguish different levels of ranking bias best when the overall distribution of advantaged and disadvantaged viewpoints is balanced. A weakness of nDJS is that its normalization causes its outcome values to be much lower in general compared to binomial fairness metrics. We note that nDJS cannot be directly compared to nDD, nDR, or nDKL and recommend to interpret nDJS carefully when ranking bias is mild.

\subsection{Caveats and Limitations}

We note that our simulation study is limited in at least three important ways. First, we consider a scenario in which documents have correctly been assigned multinomial viewpoint labels. This allows us to study their behavior in a controlled setting. In reality, existing viewpoint labeling methods are prone to biases and issues of accuracy. Current opinion mining techniques are still limited in their ability to assign such labels \cite{Wang2019} and crowdsourcing viewpoint annotations from human annotators can be costly and also prone to biases and variance \cite{Wauthier2011}.

Second, we assume that any document in a search result ranking can be assigned some viewpoint label concerning a given disputed topic. It is realistically possible for a document to contain several, or even all available viewpoints (e.g., a debate forum page). In these cases, assigning an overarching viewpoint label might oversimplify the nuances in viewpoints that exist \emph{within} rankings and thereby not leading to a skewed assessment of viewpoint diversity in the search result ranking. Future work could look into best practices of assigning viewpoint labels to documents.

Third, our simulation of multinomial viewpoint fairness included only one specific case in which one viewpoint is treated differently compared to the other six. There are other scenarios where multinomial viewpoint fairness could become relevant. These scenarios differ in how many viewpoint categories there are, how many items are advantaged in the ranking, and to what degree. Simulating all of these potential scenarios is beyond the scope of this paper. Future work could however explore how metrics such as nDJS behave in such scenarios.

\section{Conclusion}
We adapted existing ranking fairness metrics to measure binomial viewpoint fairness and proposed a novel metric that evaluates multinomial viewpoint fairness. We find that despite some limitations, the metrics reliably detect viewpoint diversity in search results in our controlled scenarios. Crucially, our simulations show how these metrics can be interpreted and their relative strengths. 

This lays the necessary groundwork for future research to assess viewpoint diversity in \emph{actual} search results. We plan to perform such evaluations of existing web search engines concerning highly debated topics and upcoming elections. Such work would not only provide tremendous insight into the current state of viewpoint diversity in search result rankings but pave the way for a greater understanding of how search result rankings may affect public opinion.

\section*{Acknowledgements}

This activity is financed by IBM and the Allowance for Top Consortia for Knowledge and Innovation (TKI’s) of the Dutch ministry of economic affairs.

We also thank Agathe Balayn, Shabnam Najafian, Oana Inel, and Mesut Kaya for their comments on an earlier draft of this paper.


\bibliographystyle{abbrv}
\bibliography{library.bib}

\end{document}